# Flux free growth of large FeSe$_{1/2}$Te$_{1/2}$ superconducting single crystals by an easy high temperature melt and slow cooling method

P. K Maheshwari[1,2], Rajveer Jha[1], Bhasker Gahtori[1], and V.P.S. Awana[1*]

[1]CSIR-National Physical Laboratory, Dr. K. S. Krishnan Marg, New Delhi-110012, India

[2]AcSIR-National Physical Laboratory, New Delhi-110012, India

ABSTRACT

We report successful growth of flux free large single crystals of superconducting FeSe$_{1/2}$Te$_{1/2}$ with typical dimensions of up to few cm. The AC and DC magnetic measurements revealed the superconducting transition temperature (T$_c$) value of around 11.5K and the isothermal MH showed typical type-II superconducting behavior. The lower critical field (H$_{c1}$) being estimated by measuring the low field isothermal magnetization in superconducting regime is found to be above 200Oe at 0K. The temperature dependent electrical resistivity ρ(T) showed the T$_c$ (onset) to be 14K and the T$_c$(ρ=0) at 11.5K. The electrical resistivity under various magnetic fields i.e., ρ(T)H for H//ab and H//c demonstrated the difference in the width of T$_c$ with applied field of 14Tesla to be nearly 2K, confirming the anisotropic nature of superconductivity. The upper critical and irreversibility fields at absolute zero temperature i.e., H$_{c2}$(0) and H$_{irr}$(0) being determined by the conventional one-band Werthamer–Helfand–Hohenberg (WHH) equation for the criteria of normal state resistivity (ρ$_n$) falling to 90% (onset), and 10% (offset) is 76.9Tesla, and 37.45Tesla respectively, for H//c and 135.4Tesla, and 71.41Tesla respectively, for H//ab. The coherence length at the zero temperature is estimated to be above 20Å by using the Ginsburg-Landau theory. The activation energy for the FeSe$_{1/2}$Te$_{1/2}$ in both directions H//c and H//ab is determined by using Thermally Activation Flux Flow (TAFF) model.

*Corresponding Author
Dr. V. P. S. Awana,
Principal Scientist
E-mail: awana@mail.npindia.org
Ph. +91-11-45609357, Fax-+91-11-45609310
Homepage awanavps.wenbs.com

INTRODUCTION

One of the most surprising discoveries in field of experimental condensed matter physics in last decade had been the observation of superconductivity in Fe based $REO_{1-x}F_xFeAs$ pnictide compounds [1-3]. Subsequently superconductivity was found in various Fe based chalcegonides as well viz. pure [4], doped [5,6] and intercalated [7-9] FeSe. The superconducting transition temperature ($T_c$) of Fe based pnictides and chalcegonides is reported in excess of above 50K, which keeps them in tune with exotic high $T_c$ cuprate superconductors [10], i.e., outside the popular BCS limit. Any reasonable and widely acceptable theoretical explanation for superconductivity above 40K (BCS strong coupling limit; the $MgB_2$ case, ref. 11) has been elusive till date. Though there are several thousand experimental research articles yet available for HTSc cuprates and Fe based superconductors, a unified one theory is yet not seen around. Clearly the superconductivity of HTSc cuprates and Fe based new superconductors is real puzzle for the theoretical condensed matter physicists.

As far as the experimental results are concerned, one always aspires for the clean single crystal data on physical properties of any new material, which is true for the exotic Fe based superconductors as well. The clean singly crystal data are a real feast for the theoreticians to work out the model basis for the observed physical properties of the given material. There are several standard techniques for obtaining single crystals of various functional materials including for superconductors. Particularly, the single crystals of Fe based chalcegonide superconductors are grown using mainly the Bridgman technique [12-14]. Basically, the constituent stoichiometric material along with melting flux (KCl in general) is melted at high temperature and subsequently cooled slowly to room temperature to obtain the desired tiny crystals. Both horizontal and vertical holding of the charge is possible in state of art relatively expensive melt furnaces. Often rotation/spinning of the melt is also desired. In, brief the crystal growth itself is not only an expensive affair but is rather state of art and an independent research field. The Fe based chalcegonide superconductors are grown by both added flux (NaCl/KCl) [15-18] and the self flux method [19-23]. Worth mentioning is the fact that the single crystals of FeSe cannot be grown directly from the melt [24]. A very recent article on growth of FeSe single crystals without flux [25] prompted us to try the same. The FeSe crystals in ref. 25 are grown without flux by travelling floating zone technique and are large enough in size for inelastic neutron scattering studies. The novel flux free growth got good appreciation [26], because this could completely avoid the contamination from foreign flux constituents if at all and also due to their relatively larger size.

We, in this short article report the successful single crystal growth of flux free $FeSe_{1/2}Te_{1/2}$ superconductor in a normal tube furnace without any complicated heating schedules related to travelling-solvent floating zone technique. The constituent stoichiometric high purity elements are mixed, vacuum sealed in quartz tube and heated to high temperature of $1000^0C$, with an intermediate step at $450^0C$ for 4hours. The hold time at $1000^0C$ is 24hours. Finally the furnace is cooled slowly (2$^0$C/minute) down to room temperature. The obtained crystals, being taken from cylindrical melt are big enough in size of around 2cmx1cmx0.5cm. Interestingly, any part taken from the melt is single crystalline, as if whole the melt is grown in crystalline form. The crystals are bulk superconducting at above 12K. The intermediate step at 450$^0$C for 4hours, while heating to melting temperature of 1000$^0$C is crucial. It seems the FeSe$_{1/2}$Te$_{1/2}$ seed is formed at this temperature, which grows in melt and gets stabilized during slow cooling to room temperature. The method thus reported is novel for obtaining single crystals of Fe chalcogenide superconductors. Worth mentioning is the fact that the method is checked for its reproducibility by several repeated runs. The flux free FeSe$_{1/2}$Te$_{1/2}$ crystals are grown earlier [19-23] by typical travelling floating zone technique applying complicated heat treatments on the other hand here we obtained the same by simple heating schedule and that also in a normal tube furnace.

EXPERIMENTAL DETAILS

The investigated FeSe$_{1/2}$Te$_{1/2}$ crystals were grown by a self flux melt growth method. The crystals had a platelet like shape and shining surfaces with typical dimensions of (2–1)cm x (1.0)cm. We took high purity (99.99%) Fe, Se and Te powder weighed them in stoichiometric ratio and grind thoroughly in the argon filled glove box. The mixed powder is subsequently pelletized by applying uniaxial stress of 100kg/cm$^2$ and then pellets were sealed in an evacuated (<10$^{-3}$ Torr) quartz tube. The sealed quartz tube is kept in automated tube furnace for heating at 450$^0$C with a rate of 2$^0$C/min for 4 h and then the temperature is increased to 1000$^0$C with a rate of 2$^0$C/min for 24h. Finally the furnace is cooled slowly with a rate of 1$^0$C/minute down to room temperature. The schematic of heat treatment is shown in Figure 1. The obtained crystals are being taken from gently crushed quartz tube. We performed room temperature x-ray diffraction (XRD)on the single crystalline material for the structural characterization using Rigaku x-ray diffractometer with CuK$_α$ radiation of 1.54184Å. The morphology of the obtained single crystal has been seen by scanning electron microscopy (SEM) images on a ZEISS-EVO MA-10 scanning electron microscope, and Energy Dispersive X-ray spectroscopy (EDAX) is employed for elemental analysis. Electrical and magnetic measurements were carried out on Quantum Design (QD) Physical Property Measurement System (PPMS) – 14Tesla down to 2K in applied fields of up to 14Tesla.

RESULTS AND DISCUSSION

Figure 2 shows the XRD pattern of FeSe$_{1/2}$Te$_{1/2}$ single crystal sample. The XRD represents only the (001), (002), (003) and (004) reflexes of tetragonal phase, which confirms the crystal growth along c lattice constant. All peaks in the X-ray diffraction pattern of the single crystals shown in Figure 2, can be attributed to tetragonal *P*4/*nmm* unit cell having a = 3.79Å and c = 5.9Å. These values are in agreement with the earlier report [27]. The photographs of the crystals are shown in Figure 3(a), clearly indicating the size of crystals to be of few centimeters. Figure 3(b) shows the typical high magnification (20μm) SEM image of as grown FeSe$_{1/2}$Te$_{1/2}$ single crystal. It is clear that the growth of the crystal is layer by layer. The scanning electron microscope image of low magnification (100μm) is shown in Figure 3(c). Though, it is a real challenge to use microscopy for proof of sample order over centimeter distances, still it is clear from Figure 3(c) that slab like layer by layer growth persists in the crystal. The slab like layer by layer growth is further shown in Figure 3(d). The compositional analysis of selected area being carried by EDX (Energy Dispersive X ray Spectroscopy) is shown in Figures 3(e) and (f), which showed that crystal, is formed in near stoichiometric composition, with slight deficiency of Se. The overall mean composition comes out to be averaged $FeSe_{0.4}Te_{0.5}$. The self flux grown $FeSe_{1-x}Te_x$ crystals were earlier reported to contain excess interstitial Fe [21-23, 28, 29]. The presently studied crystals though are Se deficient; the same do not seem to appear to have any un-reacted Fe in them and are grown by a very simple process in ordinary tube furnace.

Figure 4(a) depicts the temperature dependence of real (M') and (M") parts of AC susceptibility of $FeSe_{1/2}Te_{1/2}$ single crystal at various amplitudes in absence of dc field. It is clearly seen that a sharp decrease occurs in the real part of AC susceptibility below $T_c$, reflecting the diamagnetic shielding. In addition, below $T_c$ a sharp peak appears in M", reflecting losses related to the flux penetration inside the crystal. No indication of a two-peak behavior is detected. With increasing AC amplitudes both the volume fraction of diamagnetic shielding (M') and the peak height (M") increase monotonically. The DC susceptibility versus temperature (M-T) plot for the $FeSe_{1/2}Te_{1/2}$ single crystal is illustrated in Figure 4(b). The DC magnetization measurements are performed under applied field of 10Oe in both zero-field cooling (ZFC) and field cooling (FC) processes and the applied field is parallel to c axis of the crystal. Bulk superconductivity is confirmed as superconducting transition with an onset $T_c$ at 10.5K and an almost full shielding below 10K, whereas the field-cooled susceptibility exhibits only a small drop, possibly due to a strong flux pinning effect. More careful look of the ZFC magnetization reveals that the diamagnetic transition below $T_c$ is slightly broader and not completely saturated, indicating, possible defect and presence of weak links in studied single crystals.

Figure 4(c) shows the isothermal M-H plot for studied $FeSe_{1/2}Te_{1/2}$ single crystal at temperatures 2.2K, 5K and 25K. The magnetic hysteresis plots of $FeSe_{1/2}Te_{1/2}$ single crystal at 2.2 and 5K are evidently of a typical type-II superconductor. The M-H curve at 2.2K is wide open up to the applied field of 5Tesla, suggesting high upper critical filed for the studied $FeSe_{1/2}Te_{1/2}$ single crystal. There is evidently no ferromagnetic background in the superconducting M-H curve at 2.2 K for $FeSe_{1/2}Te_{1/2}$ single crystal. In Fe based superconductors, it is important to exclude the inclusion of un-reacted Fe impurity. One way to check this is to perform isothermal magnetization (M-H) of these compounds at just above their superconducting transition temperature ($T_c$) and check if ordered Fe or $FeO_x$ exists in the material [30,31]. The M-H for the studied $FeSe_{1/2}Te_{1/2}$ single crystal at 20K is shown in Figure 4(c). Clearly the 20K M-H is linear and without any hysteresis, thus excluding the possibility of inclusion of un-reacted ordered Fe in our crystal.

To evaluate lower critical field $H_{c1}(T)$ we measure low field M-H at different temperatures for H//ab as shown in figure 4(d). In the M-H curve the linear low-field part principally overlaps with the Meissner line due to the perfect shielding. Consequently, $H_{c1}(T)$ can be defined as the point where M-H deviates by say 2% from the perfect Meissner response. The values of $H_{c1}(T)$, thus obtained, are shown in Figure 4(e) for different temperatures. All the plots exhibit linear response for low fields ($H_{c1}$) and then deviate from linearity for higher fields. Thus determined, $H_{c1}(T)$ values are well fitted by using the formula $H_{c1}(T) = H_{c1}(0)[1-(T/T_c)^2]$, and the obtained $H_{c1}(0)$ is 204 Oe for $FeSe_{1/2}Te_{1/2}$ single crystal in H//ab condition.

Figure 5(a) shows the temperature dependence of the in-plane electrical resistivity $\rho(T)$ below 300K. The curvature of $\rho(T)$ changes at about T=150K and becomes metallic with a further decrease in temperature, superconductivity occurs with $T_c$ (onset) and $T_c(\rho=0)$ at 14K and 11.5K respectively. The temperature dependent electrical resistivity under various magnetic fields is shown in Figures 5(b) and 5(c) for both H//ab and H//c. The current was applied parallel to the ab plane in both situations. The $T_c$ (onset) and $T_c(\rho=0)$ shift towards the low temperature side with increasing magnetic fields for both field directions. Interestingly, the resistivity transition width is broader for H//c than H//ab. The shape and broadening of $\rho(T)$ for H//c is similar to 122 system [32] but relatively different from 1111 system [30,33], where it was explained by the vortex-liquid state being similar to cuprates [34]. Hence, it can be concluded that the vortex-liquid state region is narrower when sample position is H//ab. The difference in the width of superconducting transition temperature with applied fields in and out of plane is nearly 2K, indicating high anisotropy in the superconducting properties of $FeSe_{1/2}Te_{1/2}$ single crystal.

For the determination of the upper critical field ($H_{c2}$) the criteria of normal state resistivity ($\rho_n$) falling to 90% of the onset is used. The upper critical field at absolute zero temperature $H_{c2}(0)$ is determined by the conventional one-band Werthamer–Helfand–Hohenberg (WHH) equation, i.e., $H_{c2}(0)= - 0.693T_c(dH_{c2}/dT)_{T=T_c}$ for all the criteria. The estimated $H_{c2}(0)$ for 90% (onset) is 76.9Tesla for H//c and 135.4Tesla, for H//ab. In Figures 5(d) and (e), the solid lines are the extrapolation to the Ginzburg–Landau equation $H_{c2}(T)=H_{c2}(0)(1-t^2/1+t^2)$, where $t=T/T_c$ is the reduced temperature. These upper critical field value for H//ab is much higher than the H//c. The irreversible field $H_{irr}$ is determined using 10% criteria of normal state of $\rho(T)$ under various magnetic fields and is more or less half of the upper critical field. The region between irreversible field $H_{irr}(T)$ and upper critical field $H_{c2}(T)$ is liquid vortex region, which has significant importance for a superconductor. The irreversible filed $H_{irr}(T)$ for H//ab is above 70Tesla, while the $H_{irr}(T)$ for H//c is nearly 37Tesla. Clearly the superconducting response of the sample is highly anisotropic. To further determine other superconducting parameters we use the Ginzburg-Landau theory corresponding to the magneto resistivity data. The Ginzburg-Landau coherence length $\xi(0)$ is calculated by taking the values of $H_{c2}(0)$. The relation between $\xi(0)$ and $H_{c2}(0)$ is $H_{c2}(0) =\Phi_o/2\pi\xi(0)^2$ where $\Phi_o=2.0678\times10^{-15}$ Tesla-m$^2$ is the flux quantum. The coherence length at the zero temperature was estimated to be 20.6Å for H//c. On the other hand when the sample position was H//ab, the ξ(0) value is 15.5Å. The small coherence length along with the high upper critical field is clear indication of the type-II superconductivity for $FeSe_{1/2}Te_{1/2}$ single crystals. Finally, worth mentioning is the fact, that WHH formula is for single band systems and the $FeSe_{1/2}Te_{1/2}$ has been certificated as multiband and mutigap superconductor. However, in absence of any other accepted formulism, the WHH and GL are yet commonly used in case of Fe based exotic superconductors to estimate the fundamental superconducting parameters [27, 30, 32-34].

The electrical resistivity under various magnetic field of $FeSe_{1/2}Te_{1/2}$ single crystal is further investigated by using vortex glass model to estimate vortex glass state and TAFF (Thermally activated flux flow) model to calculate the activation energy. According to the vortex phase transition theory [35], for two or three dimensional systems a vortex-glass phase may occur with disappearance of resistivity and long-range phase coherence. In presence of magnetic field at T=0, the two dimensional superconductors do not show the long-range ordering. As stated in "flux-creep" models, the correlation length of the pairing field i.e., the vortex-glass phase should grow upon cooling and diverge as T→0. We suppose that the long-range ordering in the system is very similar to the magnetic order that occurs in a spin glass, the name is vortex glass phase, which can be stable at non zero temperature i.e., at the glass transition temperature ($T_g$). In the vortex glass state close to $T_g$, the resistivity disappears as a power law $\rho = \rho_0 |T/T_g - 1|^s$, where $\rho_0$ is a residual resistivity and s is a constant, both depending on the kind of disorder. Figures 6 (a) and (b) demonstrate the $(dln\rho/dT)^{-1}$ vs T in both directions H//c and H//ab of $FeSe_{1/2}Te_{1/2}$ single crystal based on the vortex glass model. The resistivity goes to zero at $T_g$ thus $T_g(B)$ can be extracted by applying the relation, $(dln\rho/dT)^{-1} \alpha (T-T_g)/s$, to the resistive tails. From resistivity power low we estimated the values of s=2.27, in the temperature range $T_g <T <T_c$. It suggests that the resistivity of two dimensional Iron based superconductor $FeSe_{1/2}Te_{1/2}$ can be described by the vortex glass model.

According to Thermal Activation Flux Flow (TAFF) theory [36, 37], the broadening in electrical resistivity with increasing magnetic field is understood with the thermally assisted flux motion. For the type-II superconductors TAFF can be lead by thermal fluctuations of vortices, from ρ(T,H) curve with increasing magnetic field the resistivity transition shifts towards lower temperature with increasing broadening. The Iron based superconductors $ReO_{1-x}F_xFeAs$ (Re-1111) showed similar transition broadening as for $YBa_2Cu_3O_7$ with increasing field, on the other hand Ba-122 compounds show negligible broadening due to low thermal fluctuations [38-40]. Interestingly, the $FeSe_{1-x}Te_x$ compounds show intermediate broadening with increasing field [5]. The resistivity in TAFF region is due to creep of vortices which is thermally activated, so that the resistivity in TAFF region of the flux creep is given by Arrhenius equation [38], i.e., $\rho(T,H) = \rho_0 \exp[-U_0/k_BT]$, where, $\rho_0$ is the temperature independent constant, $k_B$ is the Boltzmann's constant and $U_0$ is TAFF activation energy. $U_0$ depends weakly on magnetic field and orientation. The TAFF fitted (black line) electrical resistivity as $ln\rho$ vs $T^{-1}$ is shown in Figures 7(a) and (b) for H//c and H//ab. All the fitted lines cross at nearly $T_c$ i.e. 13.6K for H//c and 12.8K for H//ab. The values of the activation energy are estimated in the magnetic field range 1Tesla to 14Tesla from 65meV to 18meV for H//c and 0.5Tesla to 14 Tesla from 82meV to 25meV for H//ab. It can be seen from Figure 7c that the TAFF activation energy scales as power law ($U_0 = K \times H^{-\alpha}$) with magnetic field. Also, the field dependence of $U_0$ is different for lower and higher field values with $\alpha = 0.29$ for lower field (1-4Tesla) and $\alpha = 0.77$ for high field (6-14Tesla) for H//c, while $\alpha = 0.11$ for lower field (0.5-2Tesla) and $\alpha = 0.65$ for high field 4-14Tesla) for H//ab. The activation energy is comparatively high for H//ab than the H//c. The weak power law decrease of $U_0$ in low field for the both the field directions suggests that the single vortex pining dominates in this regimes, followed by a more rapidly decrease of $U_0$ in field, which could be related to the crossover to a collective Pinning regime [37].

CONCLUSION

We have successfully synthesized the $FeSe_{1/2}Te_{1/2}$ large (cm size) single crystals through self flux method applying a simple heating schedule in an ordinary tube furnace. The single crystal grows along the (0 0 l) plane, which has been confirmed by XRD data. The superconductivity at 11.5K has been established by both AC and DC magnetic measurements. The $\rho(T)$ measurements showed $T_c$ (onset) and $T_c(\rho=0)$ at14 K and 11.5 K respectively. The $\rho(T)H$ for H//ab and H//c showed strong anisotropy. $H_{c2}(0)$ is determined by the conventional one-band WHH equation with 90% of $\rho_n$ criterion and found to be 76.9Tesla and 135.4Tesla for H//c and H//ab respectively. Similarly, the $H_{irr}(0)$ being determined from 10% of $\rho_n$ criterion is found to be 37.45 Tesla and 71.41 Tesla for H//c and H//ab respectively. The estimated activation energy $U_o(H)$ showed weak power law decreases low fields (1-4Tesla) for the both field directions suggesting that the single vortex pining dominates in this region. The large (cm size) single crystals, which are bulk superconducting at above 12K could be good candidates for neutron scattering studies and thus to unearth the physics of these novel superconductors.

ACKNOWLEDGEMENT

Authors would like to thank their Director NPL India for his keen interest in the present work. This work is financially supported by *DAE-SRC* outstanding investigator award scheme on search for new superconductors. P. K. Maheshwari thanks CSIR, India for research fellowship and AcSIR-NPL for Ph.D. registration.

**Figure Captions**

**Figure 1:** Schematic diagram of the heat treatment used to grow $FeSe_{1/2}Te_{1/2}$ single crystal through self flux method.

**Figure 2:** XRD patterns of $FeSe_{1/2}Te_{1/2}$ single crystal at room temperature.

**Figure 3:** (a) Photograph of $FeSe_{1/2}Te_{1/2}$ single crystals (b-d) SEM images of $FeSe_{1/2}Te_{1/2}$ single crystal for 20μm, 100μm and 40μm magnification (e-f) The EDX quantitative analysis graph of the $FeSe_{1/2}Te_{1/2}$ single crystal.

**Figure 4:** (a) The AC magnetic susceptibility in real ($M'$) and imaginary($M''$) situations at fixed frequency of 333 Hz in varying amplitudes of 1–15Oe for $FeSe_{1/2}Te_{1/2}$ single crystal. (b) DC magnetization (both ZFC and FC) plots for $FeSe_{1/2}Te_{1/2}$ single crystal measured in the applied magnetic field, H = 10Oe. (c) Isothermal MH curve at 2.2K, 5K and 25K of $FeSe_{1/2}Te_{1/2}$ single crystal. (d) Low field M-H curve at 2.2K-8K $FeSe_{1/2}Te_{1/2}$ single crystal. (e) Temperature dependence of $H_{c1}(T)$, the solid with line is fitting to $H_{c1}(T) = H_{c1}(0)[1 − (T/T_c)^2]$ for $FeSe_{1/2}Te_{1/2}$ single crystal.

**Figure 5:** (a) The temperature dependent electrical resistivity in temperature range 300-5K for $FeSe_{1/2}Te_{1/2}$ single crystal. Temperature dependence of the resistivity $\rho(T)$ under various magnetic fields up to 14Tesla for (b) H//c and (c) H//ab plan for $FeSe_{1/2}Te_{1/2}$ single crystal. (d, e) The upper critical ($H_{c2}$) and irreversibility ($H_{irr}$) fields estimated from the $\rho(T)H$ data with 90%, and 10% $\rho_n$ criteria for $FeSe_{1/2}Te_{1/2}$ single crystal.

**Figure 6:** $(d\ln\rho/dT)^{-1}$ vs T to determine the vortex glass transition temperature (a) for H//c and (b) for H//ab plan of $FeSe_{1/2}Te_{1/2}$ single crystal.

**Figure 7:** $\ln\rho(T,H)$ vs 1/T in different magnetic fields (a) H//c and (b) H//ab plan for $FeSe_{1/2}Te_{1/2}$ single crystal corresponding solid line are fitting of Arrhenius relation. (c) The field dependent of Activation energy $U_o(H)$ with solid lines fitting of $U_o(H) \sim H^{-\alpha}$.

Fig. 1

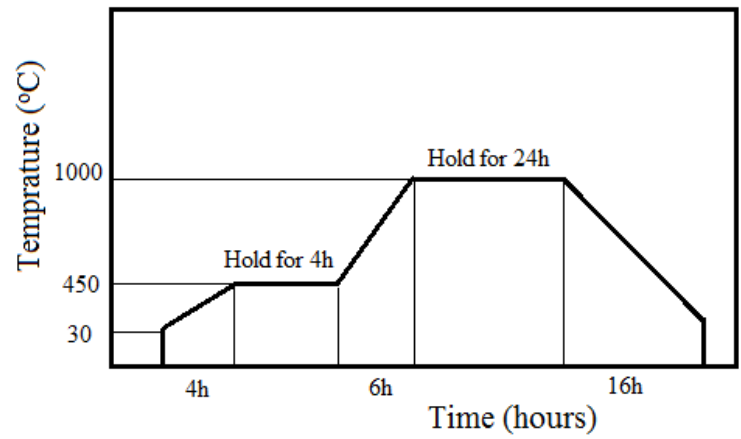

Fig. 2

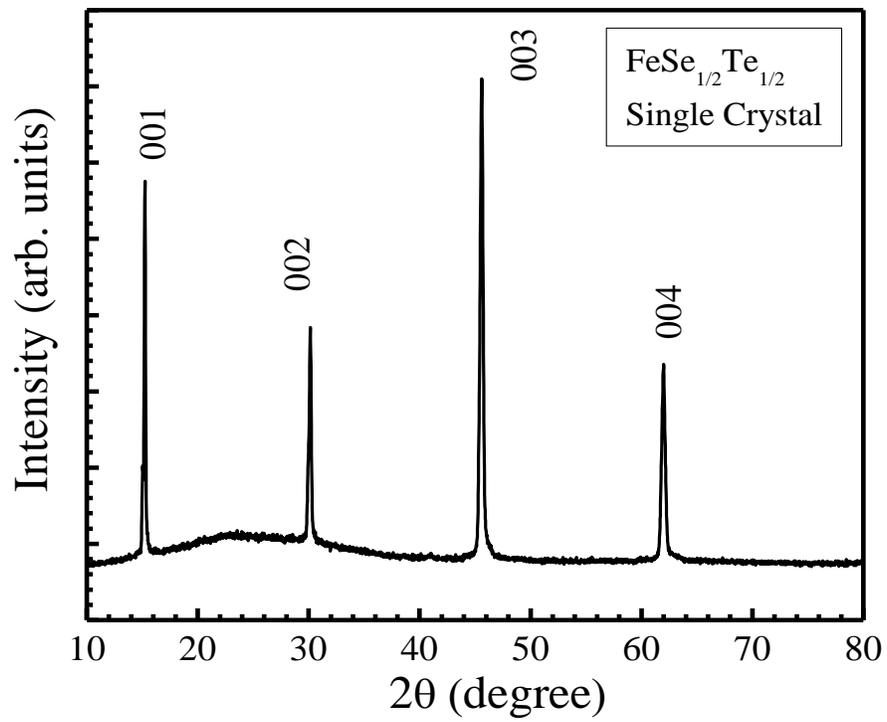

Fig. 3(a)

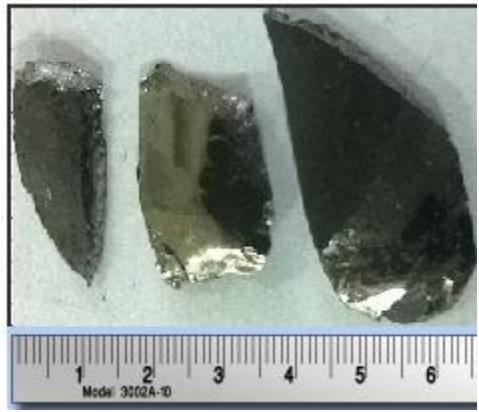

Fig. 3(b) Fig. 3(c)

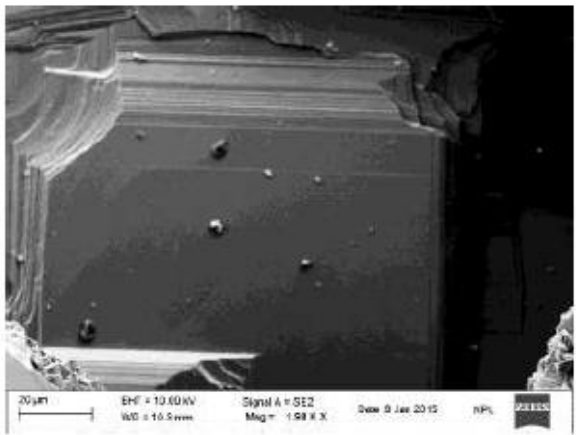 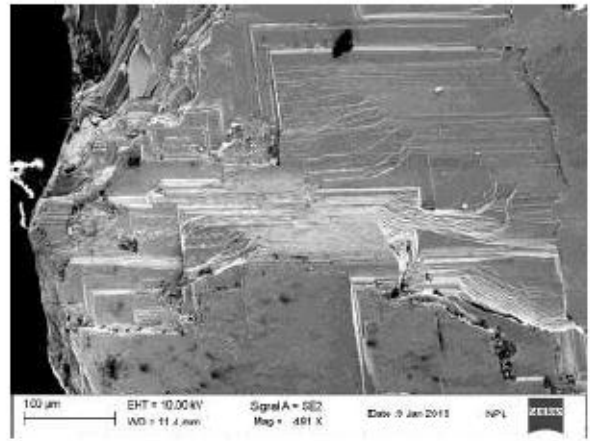

(SEM micrograph: resolution 20μm)  (SEM micrograph: resolution 100μm)

Fig. 3(d) Fig. 3(e) Fig. 3(f)

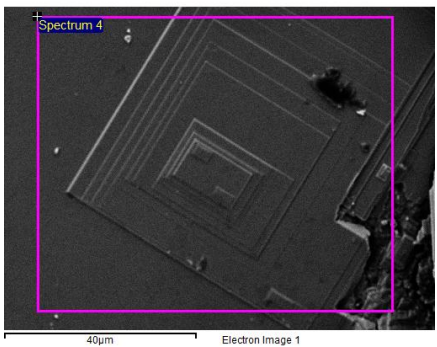 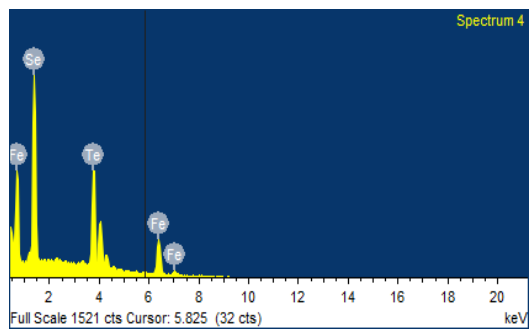 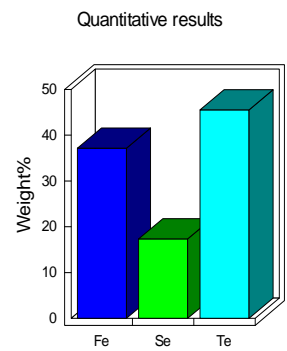

Fig. 4(a)

Fig. 4(b)

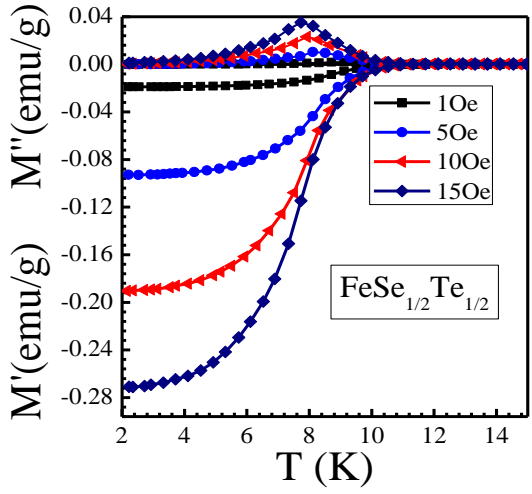
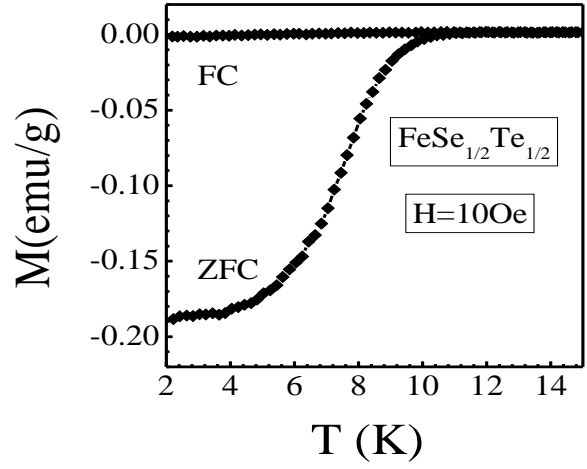

Fig. 4(c)

Fig. 4(d)

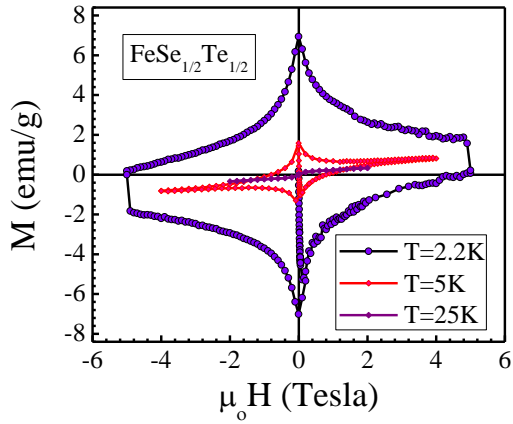
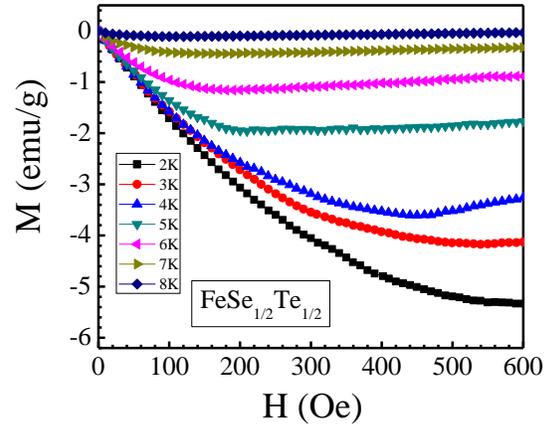

Fig. 4(e)

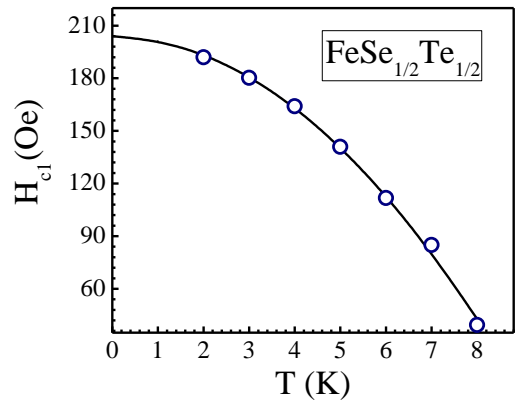

Fig. 5(a)

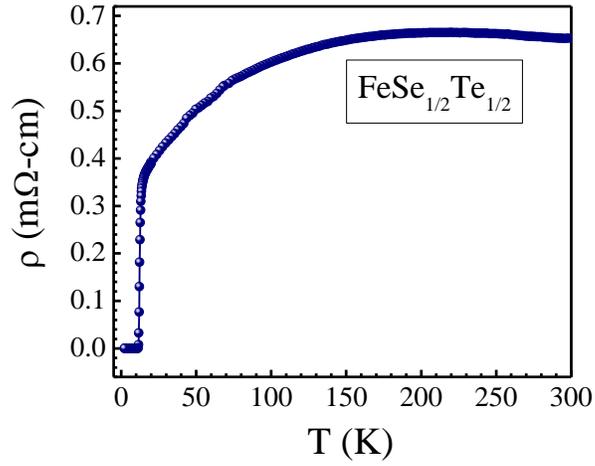

Fig. 5(b)

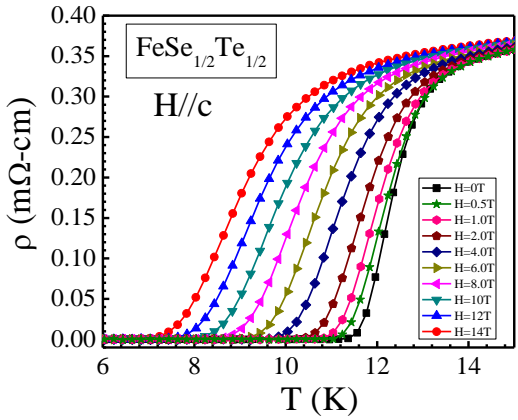

Fig. 5(c)

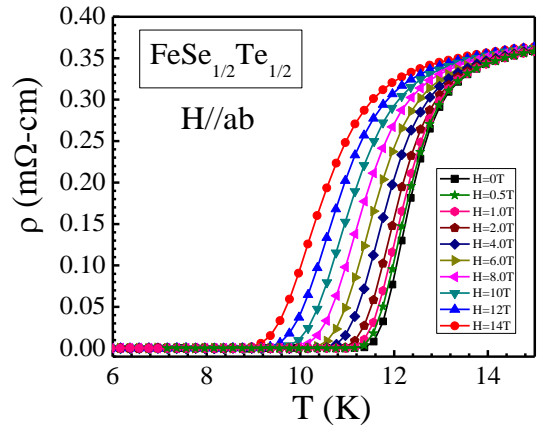

Fig. 5(d)

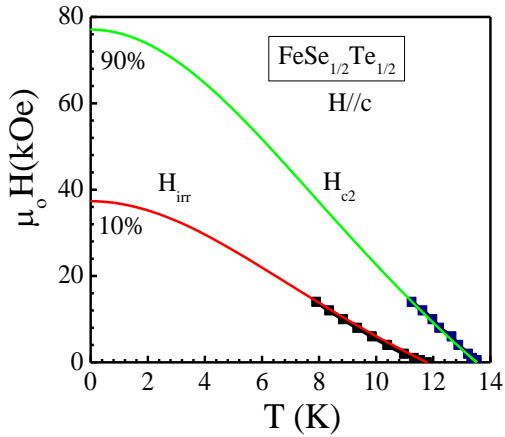

Fig. 5(e)

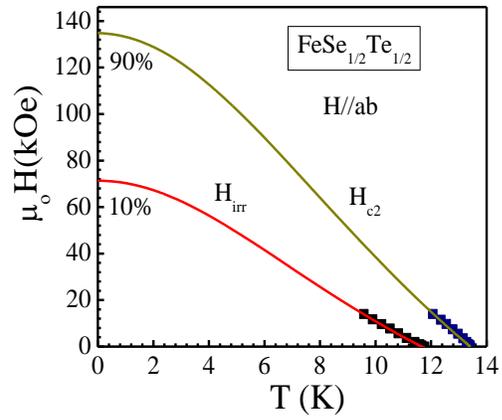

Fig. 6(a)

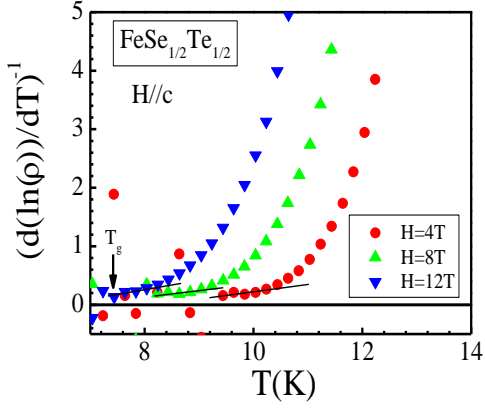

Fig. 6(b)

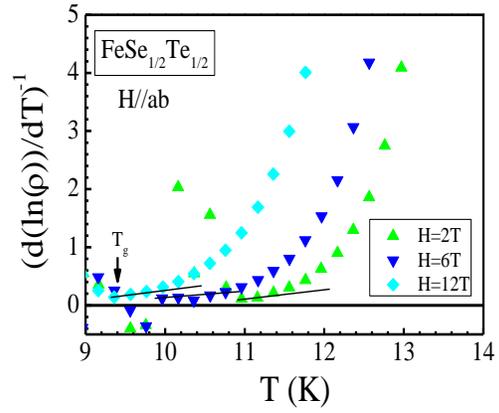

Fig. 7(a)

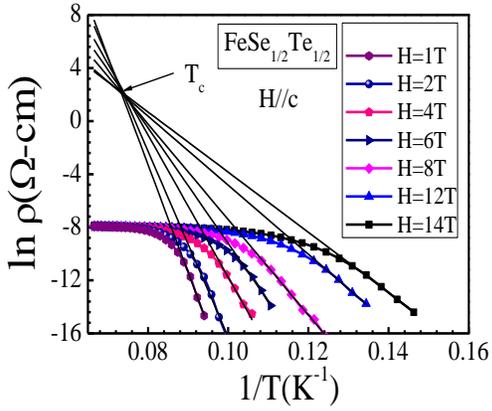

Fig. 7(b)

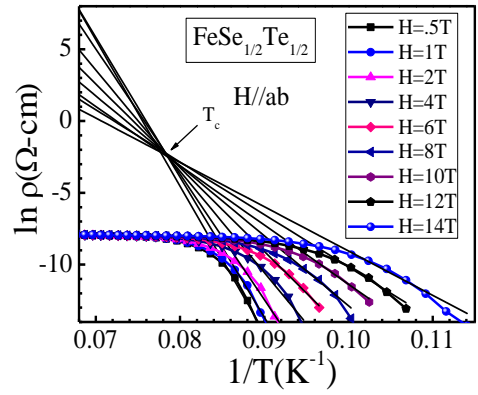

Fig. 7(c)

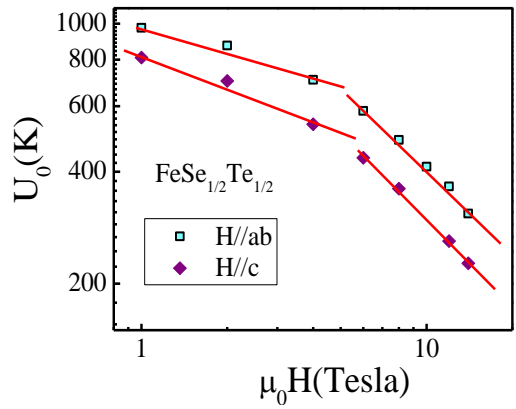